\documentclass[%
 aip,
 amsmath,amssymb,
 reprint,%
]{revtex4-1}
\usepackage{graphicx}
\graphicspath{ {./pictures/} }
\usepackage{dcolumn}
\usepackage{bm}
\usepackage{tabularx}
\usepackage[utf8]{inputenc}
\usepackage[T1]{fontenc}
\usepackage{mathptmx}
\usepackage{etoolbox}
\usepackage{tikz}
\usepackage{amsmath}
\usepackage{multirow}
\usepackage{physics}
\usepackage[version=4]{mhchem}
\usepackage{adjustbox}
\makeatletter
\def\@email#1#2{%
 \endgroup
 \patchcmd{\titleblock@produce}
  {\frontmatter@RRAPformat}
  {\frontmatter@RRAPformat{\produce@RRAP{*#1\href{mailto:#2}{#2}}}\frontmatter@RRAPformat}
  {}{}
}%

\makeatother
\begin{document}

\preprint{AIP/123-QED}

\title[]{Static Embedding with Pair Coupled Cluster Doubles Based Methods \newline}

\author{Rahul Chakraborty}
\author{Katharina Boguslawski}
 \author{Paweł Tecmer}
 
 \email{ptecmer@fizyka.umk.pl}
\affiliation{\textit{$^{1}$}~Institute of Physics, Faculty of Physics, Astronomy, and Informatics, Nicolaus Copernicus University in Toruń, Grudziądzka 5, 87-100 Torun, Poland }


\begin{abstract}

Quantum embedding methods have recently developed significantly to model large molecular structures. 
This work proposes a novel wave function theory in density functional theory (WTF-in-DFT) embedding scheme based on pair-coupled cluster doubles (pCCD)-type methods.
While pCCD can reliably describe strongly-correlated systems with mean-field-like computational cost, the large extent of dynamic correlation can be accounted for by (linearized) coupled-cluster corrections on top of the pCCD wave function.
Here we focus on the linearized coupled-cluster singles and doubles (LCCSD) ansatz for electronic ground states and its extension to excited states within the equation of motion (EOM) formalism.
We test our EOM-pCCD-LCCSD-in-DFT approach for the vertical excitation energies of the hydrogen-bonded water--ammonia complex and uranyl tetrahalides ({\ce{UO$_2$X$_4^{2-}$}, X=F, Cl, Br}).
Furthermore, we assess the quality of the embedding potential using an orbital entanglement and correlation analysis.
The approximate models successfully capture changes in the excitation energies going from bare fragments to supramolecular structures and represents a promising computation model for excited states in large molecular systems. 

\end{abstract}
\maketitle 

\section{Introduction}
The applicability of traditional highly accurate wave function theory (WFT)-based quantum chemistry methods like configuration interaction,~\cite{szalay2012} coupled cluster,~\cite{bartlett_2007} and their various truncation schemes are restricted due to their steep computational scaling. 
An alternate approach to circumvent this problem is embedding methods.~\cite{gomes_rev_2012, knizia-dm-emb-prl-2012, pernal-rdm-emb-pccp-2016}
Guided by the fact that electron correlation is \textit{`local'} in nature,~\cite{pulay1983localizability, hampel1996local} more accurate,  albeit computationally expensive higher-level methods are used to treat only a small subset of atoms in the whole molecular structure, also called the system. 
The remaining larger part of the molecular structure, often called the environment, is approximately represented by lower-level methods.~\cite{gomes_crystal, neugebauer2009subsystem, neugebauer2010chromophore, solovyeva2014describing, jacob-neugebauer-wires-review-2014} 
Hence, any computational approach based on the fragmentation~\cite{li2007generalized, dahlke2007electrostatically, pruitt2012fragment, masoumifeshani2021symmetrized} of the supramolecular structure and mix-and-match of various methods for different fragments can be regarded as embedding.
The overall goal is to reduce the computational cost of treating the supramolecule with expensive methods and focus on a small region of interest where the physical or chemical processes are centered, like electronic transitions. 

The different flavors of quantum embedding arise from combining various computational methods.
The quantum mechanics/molecular mechanics (QM/MM) variant is a popular class of embedding schemes.~\cite{lin2007qm, senn2009qm}
Since the quantum mechanical description of the environment fragment must explicitly include the purely quantum mechanical interaction terms such as Pauli repulsion, QM/QM embedding methods were introduced.
In this class of embedding approaches, the system part is described with orbitals. At the same time, the environment is mimicked only through its nuclear electrostatic potential and its electron density $\rho_{\textrm{env}}$, calculated in the presence of the electron density of the system $\rho_{\textrm{sys}}$.  

One of the most prominent examples of the QM/QM embedding approach is DFT-in-DFT embedding.~\cite{cortona_91, cortona_92, cortona-ijqc-1994, cortona-mgo-prb-1999, lda_0,lda_1}
The frozen density embedding (FDE) method was developed in this context.~\cite{fde_0, fde_1} 
In this approach, the density of the environment is kept frozen, and the system's ground state density and energy are calculated using the frozen environment density as an external field.
An iterative `freeze-and-thaw' cycle was also employed to improve the fragment (system and environment) densities in response to changes in one another.~\cite{fde-freez-and-thaw}
FDE has also been extended to excited states using time-dependent density functional theory (TD-DFT).~\cite{fde_tddft, fde_coupled, neugebauer-solvent-screening-tddft-2010} 
Still, the DFT-in-DFT approach can only give an approximate picture of the interactions between the structure fragments mainly because of the non-additive kinetic potential term present in the interaction potential and intrinsic limitations of DFT.~\cite{wesolowski-emb-review}
Developing more accurate embedding schemes is still an active field of research.
Examples are the studies with inverted potentials~\cite{wesolowki-non-additive-kin-pot-ijqc-2018, wesolowski-2022-nonadditive} and with orthogonal orbitals from different fragments.~\cite{manby-projected-embedding-2012, culpitt-emb-with-orthogonality-contrains}

Parallel to the popularity of DFT-in-DFT embedding, WFTs have also been augmented with DFT in the so-called WFT-in-DFT embedding.\cite{wftindft_1, wftindft_2, wftindft_3, gomes_crystal, wesolowski-orbital-free-embedding-pra-2008, neugebauer-densities-wfs-cpc-2016, kallay-exact-emb-jcp-2016, kallay-exact-emb-jctc-2018, embedding-review-gagliardi}
In the basic WFT-in-DFT embedding formalism, the energy functional is expressed as~\cite{wftindft_1, wftindft_2, wftindft_3,  wftindft_surface} 
    \begin{align}
     \label{eq:wft-in-dft}
E[\Psi_{\textrm{sys}}^{\textrm{WFT}},\rho_{\textrm{env}}^{\textrm{DFT}}] &= E_{\textrm{sys}}[\Psi_{\textrm{sys}}^{\textrm{WFT}}]
    +E_{\textrm{env}}[\rho_{\textrm{env}}^{\textrm{DFT}}] \nonumber\\
    &+E_{\textrm{int}}[\rho_{\textrm{sys}}^{\textrm{WFT}},\rho_{\textrm{env}}^{\textrm{DFT}}],
    \end{align}
where $E_{\textrm{sys}}$ is the energy of the embedded system calculated with WFT through its characteristic wave function $\Psi$$_{\textrm{sys}}^{\textrm{WFT}}$, whereas $E_{\textrm{env}}$ is the energy of the environment obtained with DFT using its electron density $\rho_{\textrm{env}}^{\textrm{DFT}}$. The third term in Eq.~\eqref{eq:wft-in-dft} is described as  
     \begin{align}
     \label{eq:e_int}
    E_{\textrm{int}} &= E_{\rm NN}+\int  \rho_{\textrm{sys}}(r)\nu_{\textrm{env}}^{\textrm{nuc}}dr + \int  \rho_{\textrm{env}}(r)\nu_{\textrm{sys}}^{\textrm{nuc}}dr \nonumber \\
    &+ \frac{1}{2}\int\frac{\rho_{\textrm{sys}}(r)\rho_{\textrm{env}}(r')}{|r-r'|}dr dr' \nonumber \\ 
    &+ E_{\textrm{xc}}^{\textrm{nadd}}[\rho_{\textrm{sys}},\rho_{\textrm{env}}]+ T_{s}^{\textrm{nadd}}[\rho_{\textrm{sys}},\rho_{\textrm{env}}].
 \end{align}
In the above equation, $E_{\rm NN}$ is the nuclear--nuclear repulsion energy, $\rho_{\textrm{sys}}^{\textrm{WFT}}$ is the electron density of the system fragment obtained using the WFT method of choice, $\nu_{\textrm{sys}}^{\textrm{nuc}}$ and $\nu_{\textrm{env}}^{\textrm{nuc}}$ are nuclear electrostatic potential terms from the system and environment fragments respectively.
$E_{\textrm{xc}}^{\textrm{nadd}}$ and $T_{s}^{\textrm{nadd}}$ correspond to the non-additive parts of the exchange--correlation (xc) and kinetic energy functionals, respectively.
The main attractive feature of WFT-in-DFT is that any WFT ansatz can be used along with DFT, and existing quantum chemistry codes can be modified to account for the DFT environment.~\cite{gomes_crystal, gomes-prl-embedding-2018}
The approach has drawn attention mainly due to the shortcomings of DFT in treating bond-breaking, strongly-correlated systems, and electronically excited states.
The method has seen extensive applications in surface chemistry and materials design through the pioneering works of Carter and coworkers.~\cite{wftindft_1, huang-emb-potential-2011, carter-projected-emb-implementation-2015}
Manby and coworkers did extensive work coupling DFT with correlated wave function methods like coupled cluster and other projection-based wave functions.~\cite{manby-projected-embedding-2012, manby-projecte-emb-open-shells-jcp-2012, manby-pojected-emb-jcp-2014, manby-projected-emb-multiscale-jctc-2016}
The WFT-in-DFT embedding has also been used to study reaction pathways~\cite{manby-projected-emb-multiscale-jctc-2016, manby-projected-emb-rsoc-2018} and excited state,~\cite{khait2010embedding, gomes_rev_2012, bennie2017pushing, ghosh2018combining, hofener-fde-limits-jcp-2022} including four-component Hamiltonians.~\cite{gomes_crystal} 

Given the ever-expanding horizon of electronic structure theory, there is still room for new WFT-in-DFT embedding schemes in terms of applicability and computational cost.
In this work, we propose the recently developed geminal-motivated~\cite{pawel-pccp-geminal-review-2022} pair coupled cluster doubles (pCCD) ansatz~\cite{limacher_2013,oo-ap1rog,tamar-pccd} as the WFT component in the WFT-in-DFT scheme.
Unlike any other WFT method used within the WFT-in-DFT framework, pCCD can work within its optimized orbital basis without defining active spaces. 
The orbital-optimized variant of pCCD has been shown to model strongly-correlated systems, albeit with mean-field-like computational costs effectively.~\cite{oo-ap1rog, ps2-ap1rog, tamar-pccd, piotrus_mol-phys, ap1rog-jctc} 
The size-extensive and size-consistent nature of orbital-optimized pCCD has led to a wide range of applications for covalent molecules,~\cite{pawel_jpca_2014, ps2-ap1rog, ap1rog-jctc, pawel-pccp-2015, garza2015actinide, pybest-paper, ola-tcc, ola-qit-actinides-pccp-2022} non-covalent systems,~\cite{filip-jctc-2019, garza-pccp-wdv} and excited states.~\cite{eom_pccd,eom-pccd-lccsd, pawel-yb2, pccd-ee-f0-actinides, delaram-pani-pccd-2023}
The successes of pCCD-based methods in modeling complex electronic structures motivate us to use them within the WFT-in-DFT framework.
An additional advantage of pCCD-type methods is the ability to compute orbital entanglement and correlation in terms of the single-orbital entropy and mutual information.~\cite{rissler2006, entanglement_letter, ijqc-eratum, pccd-prb-2016, post-pccd-entanglement} 
Such measures can be used to partition the supramolecule into a system and environment or assess the quality of the embedding potential in the WFT-in-DFT approach.

This work is organized as follows.
Section~\ref{sec:theory} summarizes the pCCD-based methods applied in this work, the implementation details of the employed embedding schemes, and orbital correlations.
Computation details are provided in Section~\ref{sec:comput_details}.
Section~\ref{sec:results} includes the analysis of electronic excitation energies in selected test systems obtained from supramolecular and embedding calculations.
Furthermore, we analyze the single-orbital entropies in uranyl complexes to validate the quality of the embedding potential. 
We conclude in Section~\ref{sec:conclusions}. 
\section{Theory}\label{sec:theory}
\subsection{pCCD-based methods}
One of the newest simplified members in the group of coupled cluster
methods is the pCCD ansatz~\cite{limacher_2013, tamar-pccd, pawel-pccp-geminal-review-2022} and reads

\begin{align}
     \label{eqn:pccd}
     \ket{\Psi_{\textrm{pCCD}}} = 
        \textrm{exp}\left(\sum_{i}^{\rm occ}\sum_{a}^{\rm virt}c_{i}^{a}\hat{a}_{{a}}^{\dagger}\hat{a}_{\bar{a}}^{\dagger}\hat{a}_{{\bar{i}}}\hat{a}_{i}\right)\ket {\Phi_{0}} 
        = \textrm{exp}(\hat{T}_{\rm p})\ket {\Phi_{0}},
\end{align}
where only electron-pair excitations $\hat T_{\rm p}$ enter the cluster operator, which substitutes an ($\alpha$--$\beta$) electron-pair from an occupied orbital to a virtual orbital ($i \bar i \rightarrow a \bar a$).
In the above equation, $\hat a^\dagger (\hat a)$ are the electron creation (annihilation) operators running over all virtual (occupied) orbitals $a$ ($i$).
The ${\Phi_{0}}$ is some independent particle model, usually the Hartree--Fock wave function. 
While the method is size-extensive, size-consistency is achieved through variational orbital optimization.~\cite{oo-ap1rog, ps2-ap1rog, ap1rog-jctc} 
Numerous computational studies demonstrate~\cite{tamar-pccd} that the pCCD wave function captures static electron correlation with mean field-like cost, accurately reproducing the electron correlation spectrum compared to more expensive wave-function-based methods.~\cite{pccd-prb-2016}
As pCCD includes only electron-pair excitations, it, however, misses a large fraction of the weak/dynamic electron correlation effects. 
To compensate for that, various a posteriori corrections have been proposed.~\cite{pawel-pccp-geminal-review-2022}
One example is a linearized coupled cluster (LCC) ansatz on top of the pCCD reference state.~\cite{ap1rog-lcc} 
In the LCC correction, we approximate the exponential coupled cluster ansatz with a pCCD reference wave function as
\begin{equation}
\label{eqn:lcc}
{\Psi_{\rm pCCD-LCC}}   \approx (1+\hat{T}') \ket{\Psi_{\rm pCCD}},
\end{equation}
where $\hat{T}'$ is the cluster operator containing various levels of excitations, excluding electron-pair excitations as they are already accounted for by the pCCD reference state.
When we include both single as well as non-pair double electron excitation operators,
\begin{equation}
\label{eqn:lccsd}
{\hat{T}'= \hat T_1 + \hat T_2'= \sum_{i}^{\rm occ} \sum_{a}^{\rm virt} t_{i}^a \hat{E}_{ai} + \frac{1}{2}\sum_{i, j}^{\rm occ} \sum_{a, b}^{\rm virt}~t_{ij}^{ab}~\hat{E}_{ai}\hat{E}_{bj} }, 
\end{equation}
where
\begin{equation}
\label{eqn:q_ai}
{\hat{E}_{ai} = a_{a}^{\dagger}a_{i} + a_{\bar{a}}^{\dagger}a_{\bar{i}}}
\end{equation}   
is the singlet excitation operator, we obtain the pCCD-LCCSD (pCCD-LCC singles and doubles) approach.
This method has shown quite robust performances in various theoretical studies,\cite{filip-jctc-2019, pawel-yb2, pccd-ptx, ola-tcc}, especially in cases involving the interplay of static and dynamic electron correlation.  

\subsection{Targeting excited states: EOM-pCCD} 

Excited states can be modeled within the pCCD framework using the equation of motion (EOM) formalism,~\cite{eomcc_1968, eomcc_1984, eom_lcc} giving various EOM-pCCD 
flavours depending on the choice of the cluster operator or EOM ansatz.~\cite{eom_pccd, eom_pccd_erratum, eom-pccd-lccsd, ip-pccd} 
Currently, there are three variants of EOM-pCCD available. 
The first one, EOM-pCCD+S, involves a posterior inclusion of single excitations in the EOM ansatz. 
Thus, the configuration--interaction-type excitation operator in the EOM treatment includes both the single and pair-doubles excitation operators.
Since the EOM-pCCD+S model treats the CC reference state differently (only electron-pair excitations) compared to the EOM formalism (both electron-pair and single excitations), it breaks size-intensivity.
Although being numerically marginal, the error can be cured by considering an a posteriori CCS calculation on top of the pCCD reference function, followed by an EOM treatment of electronic excited states. 
This approach was labeled as EOM-pCCD-CCS to highlight that a CCS correction on top of pCCD is applied.
To target general (non-pair) doubly excited states, the double excitation operator $\hat{T}_{2}$ is included in the ansatz for $\hat{T}$ in eq.~\eqref{eqn:lcc}. 
In that case, the pCCD reference wave function becomes pCCD-LCCSD, and the approach is known as EOM-pCCD-LCCSD.~\cite{eom-pccd-lccsd} 
The EOM-pCCD-LCCSD has shown promising results in modeling electronic excitations in polyenes and actinide systems.~\cite{pawel-yb2, pccd-ee-f0-actinides} 

\subsection{Implementation of static embedding and point charges}\label{sec:implementation_details} 
In the WFT-in-DFT scheme, the effect of the environment modeled through DFT embedding is accounted for as a static external potential. 
This external potential term includes the Coulomb potential of the nuclei and the electron density of the environment, along with the non-additive parts of the exchange-correlation and kinetic energy terms,~\cite{wftindft_1} 
\begin{align}
\label{eq:vemb}
    \nu^{\textrm{{emb}}}[\rho_{\textrm{sys}}^{\textrm{WFT}},\rho_{\textrm{env}}^{\textrm{DFT}}](r) &=\nu_{\textrm{env}}^{\textrm{nuc}}(r)+ \int dr' ~~\frac{\rho_{\textrm{env}}^{\textrm{DFT}}(r')}{|r-r'|}  \nonumber \\ 
    &+  \frac{\delta E_{\textrm{XC}}^{\textrm{nadd}}[\rho_{\textrm{sys}}^{\textrm{WFT}},\rho_{\textrm{env}}^{\textrm{DFT}}]}{\delta \rho_{\textrm{WFT}}}  \nonumber \\
    &+ \frac{\delta T_{s}^{\textrm{nadd}}[\rho_{\textrm{sys}}^{\textrm{WFT}},\rho_{\textrm{env}}^{\textrm{DFT}}]}{\delta \rho_{\small{\textrm{WFT}}}}.
\end{align}
This form of potential has seen extensive usage from solid state structures~\cite{wftindft_1, wftindft_2} 
to small hydrogen-bonded clusters.~\cite{desantisenvironmental2020}

In this work, we follow the standard procedure of the static embedding approach and further simplify the WFT-in-DFT embedding model by arguing that for molecules where DFT can describe the ground state electronic structure reasonably well, $\rho_{\rm sys}$ can also be obtained through DFT calculations.
Thus, the $\rho_{\textrm{sys}}^{\textrm{WFT}}$ term can be replaced by $\rho_{\textrm{sys}}^{\textrm{DFT}}$ in eq.\eqref{eq:vemb}. This assumption helps us to avoid multiple computationally expensive and non-trivial WFT calculations in the \textit{freeze-and-thaw} cycles just to obtain the particular density term.
Hence, we employ a DFT-in-DFT embedding technique to calculate $\rho_{\rm sys}$ and $\rho_{\rm env}$. 
Once the embedding potential is obtained, the matrix elements of $\nu^{\textrm{{emb}}}(r)$ represented in the atomic orbital basis set are calculated as
\begin{equation}
\label{eqn:potential}
\nu_{{ij}}=\bra{{\phi_{{i}}}}\nu^{\textrm{emb}}(r)\ket{\phi_{{j}}} \approx \sum_{{k}}w_{{k}}\nu^{\textrm{emb}}(r_{{k}})\phi_{i}(r_{{k}})\phi_{{j}}(r_{{k}}),
\end{equation}
where $i,j$ labels atomic orbitals, and the embedding potential is approximated on a grid with grid points $\{r_k\}$ and associated weights $w_k$.
Thus, the embedding potential expressed in the atomic orbital basis can be straightforwardly determined by evaluating the (product of two) atomic orbitals for each grid point weighted by the embedding potential at this point.
The resulting one-particle term is then included in the Hamiltonian of the WFT calculation. 

To account for external point charges, we compute an additional set of integrals (in the atomic orbital basis) due to the electron--external-point-charges attraction/repulsion and the potential energy term due to the Coulomb interaction between the external point charges.
To properly combine the point-charge model with the Douglas--Kroll---Hess scalar relativistic Hamiltonian (DKH), the DKH transformation also includes the external potential/integrals due to the point-charges to avoid large picture-change errors.~\cite{kello1998picture, dyall_book}
\subsection{Orbital entanglement analysis}
A quantitative measure of interactions between two orbitals can be achieved by calculating the single-orbital s(1)$_i$ and two-orbital s(2)$_{ij}$ entropies.
The first term s(1)$_i$ quantifies the interaction of the \textit{i}-th orbital with all other orbitals and reads 
\begin{equation}
\label{eqn:single-orb-entropy}
s(1)_{i}= -\sum_{\alpha=1}^4 \omega_{\alpha,i}\ln \omega_{\alpha,i},
\end{equation}
where $\omega_{\alpha,i}$ are the eigenvalues of the one-orbital reduced density matrix (1-ORDM) for the \textit{i}-th orbital. 
The 1-ORDM is determined from an $N$-particle RDM $\hat{\rho}=\ket{\Psi}\bra{\Psi}$, tracing out all degrees of freedom other than those of the \textit{i}-th orbital. 
Hence, the dimension of the 1-ORDM is equal to that of the one-orbital Fock space, which is 4 for spatial orbitals (unoccupied, doubly occupied, occupied by a single $\alpha$ or $\beta$ electron). 

Similarly, s(2)$_{ij}$ measures the interaction of a pair of orbitals $(i,j)$ with other orbitals, calculated from the eigenvalues $\omega_{\alpha,i,j}$ of the two-orbital RDM (2-ORDM), 
\begin{equation}
\label{eqn:two-orb-entropy}
         s(2)_{i,j}= -\sum_{\alpha=1}^{16} \omega_{\alpha,i,j}\ln \omega_{\alpha,i,j},
\end{equation}
where we have 16 possible occupations for a pair of spatial orbitals.
The total correlation between any pair of orbitals can be determined through the orbital-pair mutual information I$_{i|j}$, which is calculated as
\begin{equation}
\label{eqn:mutual_info}
     I_{i|j}=s(1)_{i}+s(1)_{j} - s(2)_{i,j},
\end{equation}
with the constraint $i\neq j$ to exclude self-correlations. 

These orbital-based entanglement and correlation measures have recently been applied to quantify orbital interactions~\cite{rissler2006, entanglement_letter, entanglement-jctc-2013, corinne_2015, ijqc-eratum, stein2016} in various chemical processes, like bond formation.
Since these measures are orbital-dependent and a change in the environment will translate to a variation in single-orbital entropies and the orbital-pair mutual information, they allow us to probe the change in orbital interactions/correlations with respect to a change in the environment.
Specifically, $s(1)_{i}$, $s(2)_{i,j}$, and $I_{i|j}$ (evaluated for similar orbitals) will differ for the bare subsystem and the supramolecular system. In contrast, the entanglement and correlation spectrum of the embedded system should qualitatively agree with the corresponding spectrum of the latter due to the perturbations experienced by the orbitals in the presence of the environment electron density.
Furthermore, the `better' the description of the static embedding, the closer the agreement in the exact and approximated entanglement and correlation spectra (quantitative agreement).
Such entanglement and correlation measures can thus be used as a reliable tool for selecting molecular subspaces.

\section{Computational details}\label{sec:comput_details}
\subsection{Structures}

The geometry of the \ce{H2O\cdots NH3} is optimized with the coupled cluster singles doubles and perturbative triples (CCSD(T)) method and the augmented Dunning-type correlation consistent basis sets of quadruple-$\zeta$ quality (aug-cc-pVQZ).~\cite{basis_dunning} 
Coordinates of the optimized structures are provided in Table S1 of the ESI\dag. 
The structural parameters of the uranyl tetrahalides are taken from Ruip\'{e}rez et al.~\cite{ruiperez-uo2cl4-jpca-2010} Point group symmetry is not used in any of the DFT-in-DFT or WFT-in-DFT calculations.
\subsection{Generation of the embedding potential and description of the point charge model}
The embedding potentials are generated within the Amsterdam Modeling Suite (AMS2022)~\cite{adf1, ams2022, adf2}
and then extracted with the help of the PyADF~\cite{pyadf} scripting framework. 
For the uranyl tetrahalides, the scalar ZORA Hamiltonian is employed.~\cite{zora}
In all DFT-in-DFT calculations, we used the triple-$\zeta$ double polarization (TZ2P) basis set,~\cite{adf_b} the PW91~\cite{pw91_xc, pbex} xc functional, and the PW91k~\cite{pw91k} kinetic energy functional. 
The PW91k kinetic functional has been used in a range of embedding studies~\cite{wesolowski-approx-kin-pot-hydrogen-bonds-jcp-1997, wesolowski-approx-kin-pot-vdw-jcp-1998, gotz-kin-pot-benchmark-jctc-2009, pawel3} and proved to be reliable. 
In our embedding DFT-in-DFT calculations, a supramolecular DFT calculation is done, followed by DFT calculations on the unrelaxed system and environment fragments. 
Thus, we did not explore the effect of the choice of the kinetic energy functional on our DFT-in-DFT potential. 
Then, a \textit{freeze-and-thaw} cycle is employed with an interchange of the definition of the system and environment part.
Subsystem DFT calculations are done on each of them, allowing each to relax in the presence of the electron density of the other for a fixed number of iterations.
After 20 \textit{freeze-and-thaw} cycles, the DFT-in-DFT embedding potential is extracted to perform further quantum chemical calculations in PyBEST~\cite{pybest-paper}.

Due to the partial ionic nature of the bond between the uranium atom and the halides, a point-charge embedding has also been tested as an approximate model for the halide atoms.~\cite{liu_uranyltetrachloride, gomes-uranyl-tetrachloride-emb-pccp-2013}
For that purpose, we included four negative point charges of $-0.5e$ each, centered at the halide coordinates, to replace the halide atoms in a given supramolecular geometry.

\subsection{Wave function methods}
The pCCD-based electronic structure calculations were carried out in a developer version of the PyBEST software package,~\cite{pybest-paper} while the EOM-CCSD calculations were done in the \textsc{Molpro2020.2.1} software package.~\cite{molpro-wires, molpro2020_jcp, molpro2020-authors} 
Cholesky decomposition was used for the \ce{H2O \cdots NH3} calculations in \textsc{PyBEST}.~\cite{cholesky-review-2011}
First, we performed orbital optimized pCCD~\cite{oo-ap1rog, ap1rog-jctc} calculations for the ground state and corrected for the missing dynamical correlation energy using an LCCSD correction (pCCD-LCCSD).~\cite{ap1rog-lcc} 
Pipek--Mezey orbital localization is performed ~\cite{pipek-mezey-localization} before the orbital optimization process to accelerate the convergence. 
The excitation energies were obtained within the EOM formalism~\cite{rowe-eom, bartlett-eom} on top of pCCD-LCCSD.~\cite{eom-pccd-lccsd}
In each system, we calculated the 15 lowest-lying roots.  

In our EOM-pCCD-LCCSD and EOM-CCSD calculations on the \ce{H2O \cdots NH3} complex, we applied the aug-cc-pVDZ basis set.~\cite{basis_dunning} 
For the uranyl tetrahalides, we used the ANO-RCC-VDZP basis set~\cite{basis-ano-rcc-main, roos2005_ano_actinides} and the DKH Hamiltonian of second order (DKH2).~\cite{dkh1, dkh2, dhk-hamiltonian, dkh_book} 
In correlated calculations for uranyl and its tetrahalides, we introduced a frozen core, where the 1s--4d orbitals are kept frozen for U, along with the 1s of O and F, 1s--2p of Cl, and 1s--3d orbitals of the Br atoms. 

    \begin{figure}[b]
       \centering
        \includegraphics[width=\columnwidth]{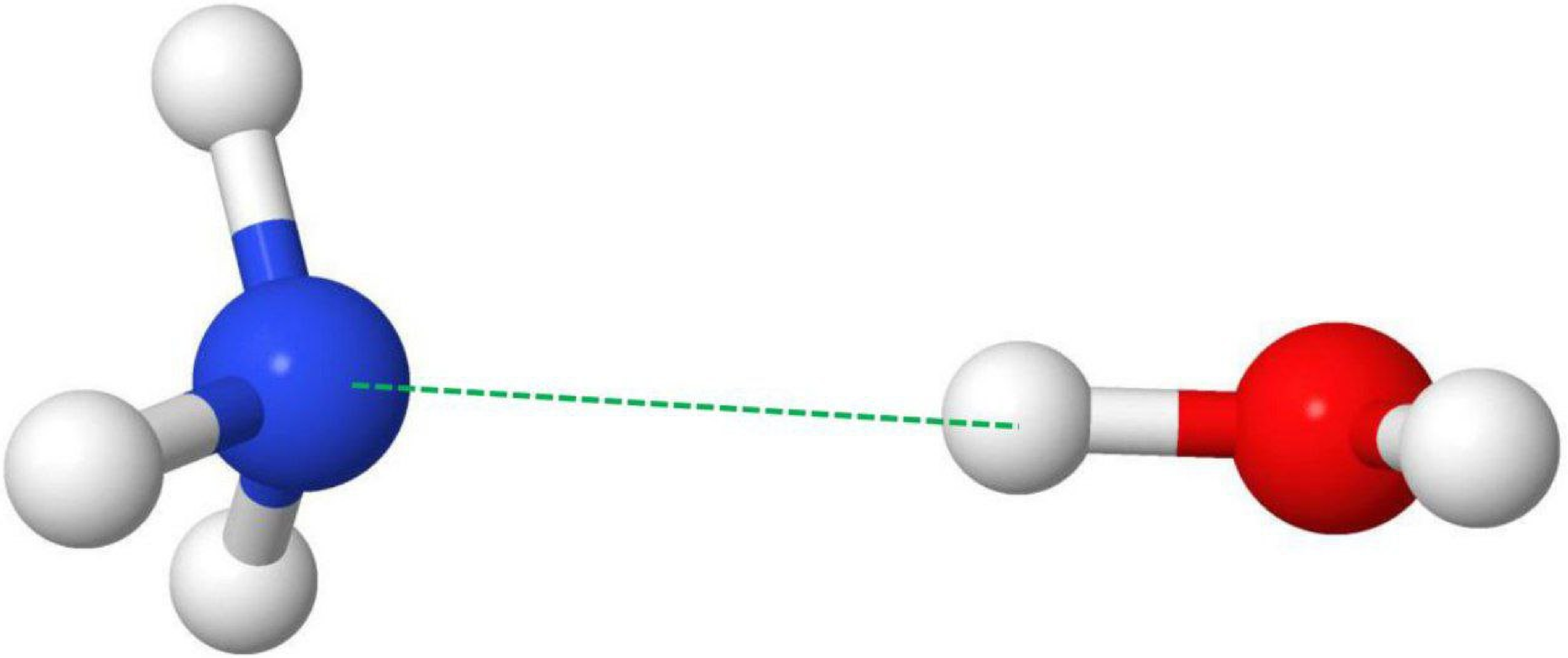}
        \caption{Structure of \ce{H2O \cdots NH3}. The hydrogen-bond is shown in green line.}
        \label{fig:water-ammonia}
    \end{figure}

\section{Results and discussion}\label{sec:results}

In the following, we assess the performance of our proposed WFT-in-DFT methods for electronic excitation energies in the model water-ammonia complex and uranyl tetrahalide complexes ({\ce{UO$_2$X$_4^{2-}$}, X=F, Cl, Br}). 
The latter are further compared to the more approximate point charge models and the supramolecular results. 
\subsection{Water-ammonia complex}
The water-ammonia complex (\ce{H2O\cdots NH3}), shown schematically in Figure \ref{fig:water-ammonia}, is a commonly used moiety for testing quantum embedding models.~\cite{hofener-wftinwft-jcp-2012,desantisenvironmental2020}
The structure can be easily partitioned into fragments that interact weakly.
The ground state of the supramolecular structure is well-represented by single-reference methods, making it a suitable test case for any embedding study. 

Table~\ref{tab:water-ammonia} lists the four lowest-lying vertical excitation energies computed with the EOM-pCCD-LCCSD method.
The fragment and supramolecular EOM-pCCD-LCCSD calculations align with the EOM-CCSD results in Table S2 of the ESI\dag.
In addition, the shifts from fragment to supramolecular structure calculated with EOM-pCCD-LCCSD are in very good agreement with  EOM-CCSD (cf. Table S3 of ESI\dag).
That observation builds upon an earlier study by some of us~\cite{pccd-ee-f0-actinides} showing that EOM-pCCD-LCCSD is, on average, closer to EOM-CCSD(T) than EOM-CCSD.
Hence, the supramolecular EOM-pCCD-LCCSD results work as a suitable reference to test our embedding approach.

The embedding calculation is set up by taking the fragment, which is the source of the excitation in the supramolecular case, as the system and the other fragment as the environment.
These are denoted as \ce{NH3} and \ce{H2O} fragments, respectively, in Table~\ref{tab:water-ammonia}. 
The \ce{NH3}-in-\ce{H2O} embedding correctly mimics the supramolecular excitation energies, and the errors are no larger than 0.15 eV. 
The \ce{H2O}-in-\ce{NH3} embedding excitation energy is underestimated by 0.18 eV compared to the supramolecular structure.
We should note that similar errors were observed in the WFT-in-DFT study by H\"ofener et al.~\cite{hofener-wftinwft-jcp-2012} 
What is important is that both the \ce{H2O}-in-\ce{NH3} and \ce{NH3}-in-\ce{H2O} embedding models predict the right signs of the supramolecular shifts with respect to their fragments. 
Finally, errors in our WFT-in-DFT approach can be mainly attributed to the limitation of the kinetic energy functional, not the method itself.

\begin{table}[b]
\small
\caption{EOM-pCCD-LCCSD vertical excitation energies (in eV) for the \ce{H2O \cdots NH3} model system.
Comparison between the bare fragments, EOM-pCCD-LCCSD-in-DFT embedding, and supramolecular EOM-pCCD-LCCSD results. The environmental shifts, from fragment to complex, calculated with WFT-in-DFT and supramolecular approaches are given in parenthesis.} 
\label{tab:water-ammonia}
\centering
\begin{tabular*}{0.49\textwidth}
{@{\extracolsep{\fill}}ccc} \hline \hline
    \multicolumn{1}{c}{Fragment} & {WFT-in-DFT}  & {supramolecule}\\ 
    &&\\\hline
\ce{NH3} & \ce{NH3}-in-\ce{H2O}&\ce{H2O \cdots NH3} \\ \hline
&&\\
6.49      &7.05 (0.56)&6.92 (0.43) \\ 
 
8.03     &8.63 (0.60)&8.49 (0.46)  \\ 
8.03    &8.67 (0.64)&8.54 (0.51) \\ 
&&\\
\ce{H2O} & \ce{H2O}-in-\ce{NH3}&\ce{H2O \cdots NH3} \\ \hline
&&\\
7.66   &7.38($-$0.28) &7.56 ($-$0.10)\\
 \hline 
 \hline

    \end{tabular*}
\end{table}

    \begin{figure}
       \centering
        \includegraphics[width=\columnwidth]{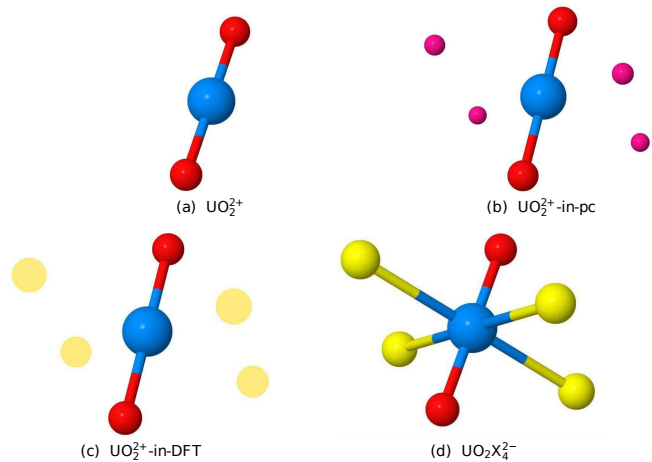}
        \caption{Schematic diagram of uranyl tetrahalide models studied}
        \label{fig:uranyls}
    \end{figure}

\subsection{Uranyl tertahalides}

\subsubsection{Analysis of electronic spectra}

The reliable modeling of the electronic structures and excitation energies of uranium-containing compounds remains a challenge for present-day quantum chemistry.
The main reasons behind it are the presence of energetically very close-lying excitations and the involvement of the f orbitals.
pCCD-based methods represent a promising alternative to standard electronic structure methods in modeling actinide chemistry.~\cite{pawel-pccp-2015, garza2015actinide, ap1rog-jctc, pccd-ee-f0-actinides, ola-qit-actinides-pccp-2022}
\begin{table*}[ht!]
\small
\caption{Vertical excitation energies (in eV) of uranyl and uranyl tetrahalides calculated with EOM-pCCD-LCCSD. The WFT-in-pc and WFT-in-DFT denote the \ce{UO$_2^{2+}$} embedded in point charges and static DFT embedding, respectively.
The transitions (except $\delta_u^{\prime}$) and molecular term symbols are labeled according to their characters in the bare uranyl. Environmental shifts from bare uranyl are shown in parenthesis. }
\label{tab:uranyls}
      \begin{tabular*}{\textwidth}{@{\extracolsep{\fill}}lllllll} \hline \hline
 
       System & Character (Term) & \ce{UO$_2^{2+}$} & {WFT-in-pc} & {WFT-in-DFT} & supramolecule \\ \hline

        \multirow{8}{*}{\ce{UO$_2$F$_{4}^{2-}$}}&$\sigma_u\rightarrow\phi_u$ ($^1\Phi_g$)&3.50 &6.32 (2.82)& 4.76 (1.26)&4.71 (1.21) \\ 
        &  & & &&  \\
        &$\sigma_u\rightarrow\delta_u$ ($^1\Delta_g$)&3.89 &4.30 (0.41)&4.38 (0.49) &4.24 (0.35)  \\ 
        &$\sigma_u\rightarrow\delta_u^{\prime}$ ($^1\Delta_g$)&3.89 &4.99 (1.10)&4.62 (0.73)&4.52 (0.63)  \\ 
        &  & & &&  \\
        &$\pi_u\rightarrow\delta_u$ ($^1\Gamma_g$)&4.82 &5.91 (1.09) &5.83 (1.01)&5.80 (0.98)  \\ 
        &$\pi_u\rightarrow\delta_u^{\prime}$ ($^1\Pi_g$)&4.94 &6.54 (1.60) &6.10 (1.26) &6.06 (1.12)   \\   
        &  & & &&  \\\hline
        
        \multirow{8}{*}{\ce{UO$_2$Cl$_{4}^{2-}$}} &$\sigma_u\rightarrow\phi_u$ ($^1\Phi_g$) &3.68 &5.14 (1.46) &4.53 (0.85) &4.48 (0.80) \\ 
        &  & & &&  \\
        &$\sigma_u\rightarrow\delta_u$ ($^1\Delta_g$) &4.05 &4.43 (0.38) &4.46 (0.41) &4.33 (0.28)  \\ 
        &$\sigma_u\rightarrow\delta_u^{\prime}$ ($^1\Delta_g$) &4.05 &4.82 (0.77) &4.57 (0.52) &4.46 (0.41) \\ 
        &  & & &&  \\
        & $\pi_u\rightarrow\delta_u$ ($^1\Gamma_g$) &5.02 &5.99 (0.97) &5.81 (0.79) &5.48 (0.46) \\ 
        &$\pi_u\rightarrow\delta_u^{\prime}$ ($^1\Pi_g$)& 5.13 &6.39 (1.26) &5.97 (0.84) &5.60 (0.47) \\
        & & & &&  \\ 
        &  & & &&  \\\hline

        \multirow{8}{*}{\ce{UO$_2$Br$_{4}^{2-}$}} &$\sigma_u\rightarrow\phi_u$ ($^1\Phi_g$) &3.69&5.04 (1.35)&4.55 (0.86)&4.24 (0.55)\\ 
        &  & & &&  \\
        &$\sigma_u\rightarrow\delta_u$ ($^1\Delta_g$) &4.07&4.43 (0.36)&4.50 (0.43)& 4.11 (0.04) \\ 
        &$\sigma_u\rightarrow\delta_u^{\prime}$ ($^1\Delta_g$) &4.07 & 4.76 (0.69)&4.59 (0.52)&4.24 (0.17) \\ 
        &  & & &&  \\
        &$\pi_u\rightarrow\delta_u$ ($^1\Gamma_g$)&5.05 &5.97 (0.92)&5.85 (0.80)& 4.76 (-0.29) \\ 
        &$\pi_u\rightarrow\delta_u^{\prime}$ ($^1\Pi_g$)&5.16  &6.32 (1.16)&6.00 (0.84) &4.84 (-0.32) \\    
        \hline 
        \hline
      \end{tabular*}
\end{table*}
Uranyl tetrahalides are the prototypes of many important actinyl complexes, and their electronic structure and photophysics have been the subject of many theoretical studies.~\cite{bursten_91, schreckenbach-actinides-1999, shamov07, schreckenbach10, matsika_cs2uo2cl4, 2-2-eom-theory, pawel2, uo2cl42-_2012,  gomes_crystal, uo2x4-eom-ccsd-ijqc-2022}
Their electronic structure properties and structural parameters are often compared to the bare uranyl cation. 
Figure \ref{fig:uranyls} schematically shows the structures of the uranyl and its tetrahalides, including the two model systems where we approximate the tetrahalide environment with point charges (Figure~\ref{fig:uranyls}b) and a static embedding potential (Figure~\ref{fig:uranyls}c).
The presence of ligand-to-metal charge transfer transitions in the upper part of the spectrum poses an additional challenge.~\cite{ruiperez-uo2cl4-jpca-2010, pawel2}
In this work, we track the changes caused by the introduction of the halide atoms in three types of the lowest-lying transitions occurring in bare uranyl and investigate whether the trends can be predicted with embedding models.

Table~\ref{tab:uranyls} shows the vertical excitation energies of the bare uranyl, uranyl with point charges (WFT-in-pc), uranyl with DFT embedding (WFT-in-DFT), and the supramolecular \ce{UO$_2$X$_4^{2-}$} systems. 
In Table~\ref{tab:uranyls}, we focus only on the lowest-lying transition energies located on the uranyl fragment.
Specifically, we analyze the excitation energies, including an electron transfer from the occupied $\sigma_u$ and $\pi_u$ orbitals to the virtual $\phi_u$ and $\delta_u$ orbitals.
In the bare uranyl, the $5f_{\phi}$ and $5f_{\delta}$ of U do not participate in the U--O bonding and hence remain purely atomic.~\cite{bursten_91, denning2007_actinyl, pawel-pccp-2015}
The lowest excitations are $\sigma_u\rightarrow\phi_u$ across the series. 
The following transitions are of $\sigma_u\rightarrow\delta_u$ nature. 
The $\phi_{u}$ and $\delta_{u}$ orbitals remain degenerate in the absence of ligands, but the $\delta_{u}$ orbitals are split in the presence of external fields. 
These are donated as $\delta_{u}$ and $\delta_{u}^{\prime}$, respectively.
The third transition is of the $\pi_u\rightarrow\delta_u$/$\pi_u\rightarrow\delta_u^{\prime}$ character. 

The U--O bond lengths in each structure are similar.
Thus, the electronic excitation energies for uranyl shown in Table~\ref{tab:uranyls} are also alike. 
The presence of the equatorial fluorine, chlorine, and bromine ligands significantly perturbs the electronic spectra of uranyl. 
The most significant perturbation is observed for the \ce{UO$_2$F$_4^{2-}$} molecule, where the effect is as substantial as 1.2 eV. 
Moving down the halogen group, the impact of the environment decreases from \ce{UO$_2$F$_4^{2-}$} to \ce{UO$_2$Cl$_4^{2-}$} to \ce{UO$_2$Br$_4^{2-}$}. 
A similar trend is observed in the EOM-CCSD excitation energies listed in Table S4 of the ESI$\dagger$. 
The perturbation effect due to the presence of equatorial ligands is reasonably well predicted by both the point-charge and static embedding models.
However, the WTF-in-DFT scheme provides more consistent energies with the reference supramolecular spectra across the \ce{UO$_2$X$_4^{2-}$} series and overall smaller errors (compare the numbers in parentheses in Table~\ref{tab:uranyls}). 
The best agreement between the WFT-in-DFT and supramolecular spectra is obtained for \ce{UO$_2$F$_4^{2-}$}.
The performance of WFT-in-DFT embedding drops down significantly for the  $\pi_u\rightarrow\delta_u$ and $\pi_u\rightarrow\delta_u^{\prime}$ electronic transitions in \ce{UO$_2$Cl$_4^{2-}$} and \ce{UO$_2$Br$_4^{2-}$}. 
For heavier halides, these electronic transitions include some admixture of ligand-to-metal charge transfer character, which is challenging to mimic within the static WFT-in-DFT approach. 

    \begin{figure*}
       \centering
        \includegraphics[width=\textwidth]{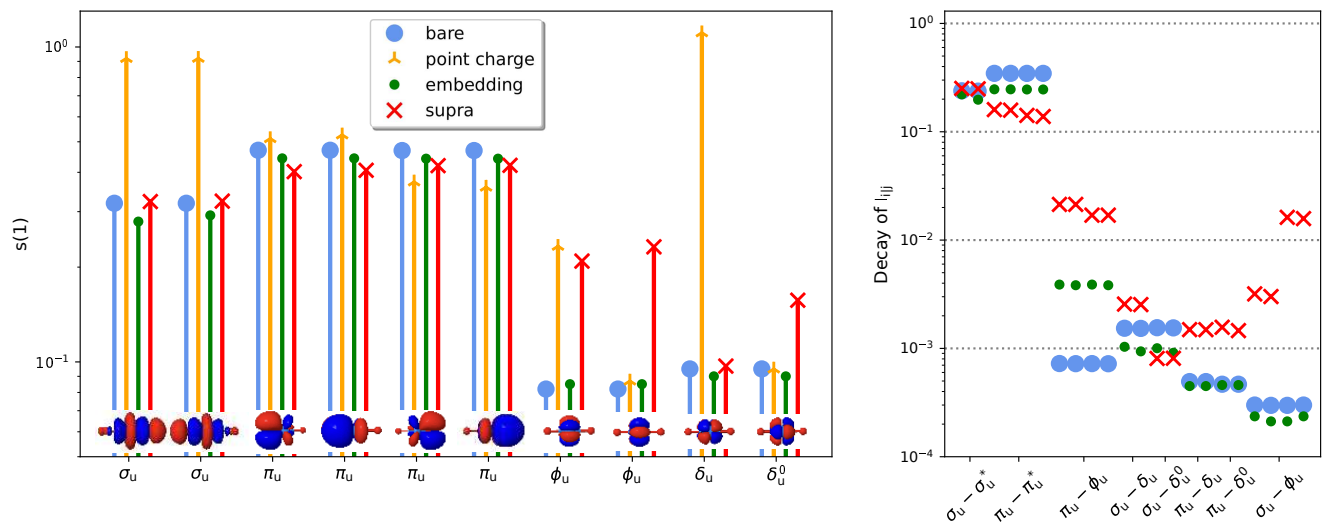}
        \caption{Single orbital entropies (left panel) and decay of the mutual information (right panel) of selected orbitals in \ce{UO$_{2}$F$_{4}^{2-}$} and approximate models. The orbitals correspond to the bare uranyl and serve as a guide for the eye. Figure S4 of the ESI$\dagger$ lists the orbitals involved in the excitation energies from different models.}
        \label{fig:s1-f}
    \end{figure*}

\subsubsection{Orbital entanglement and correlation}

    \begin{figure*}
       \centering
        \includegraphics[width=\textwidth]{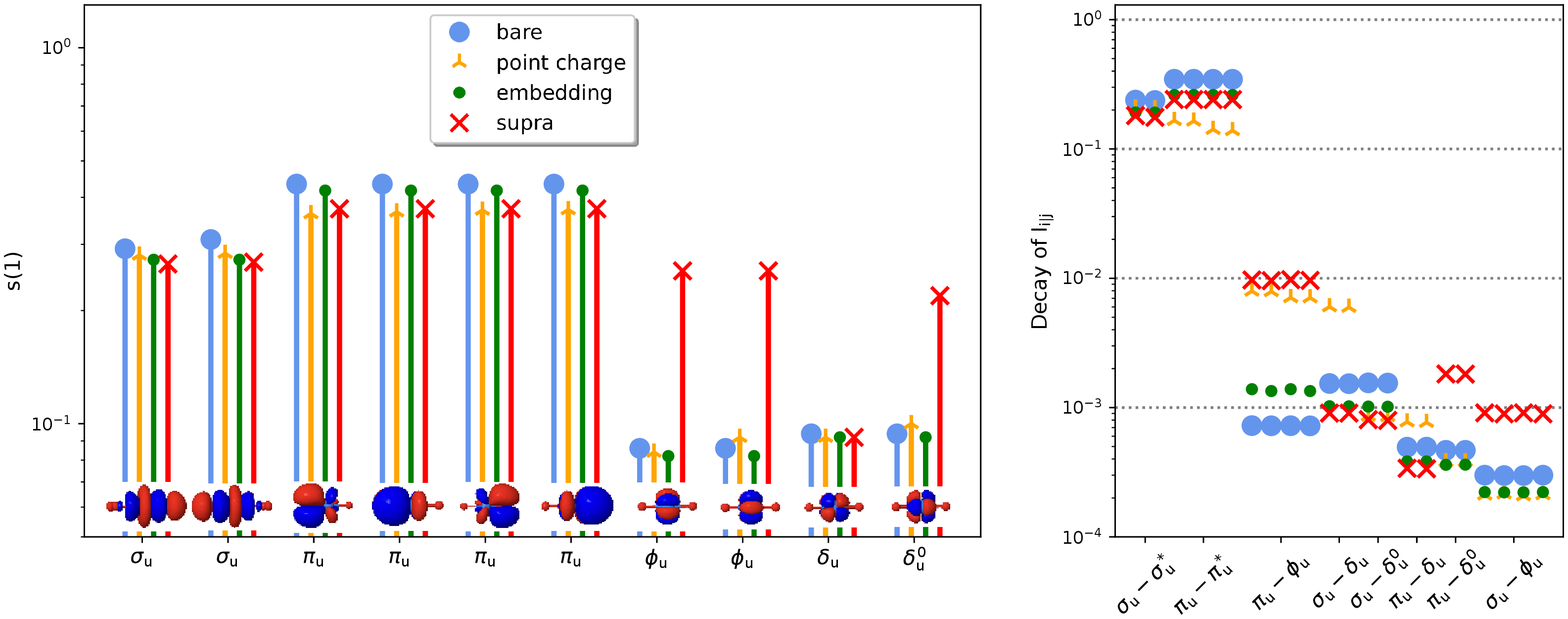}
        \caption{Single orbital entropies (left panel) and decay of the mutual information (right panel) of selected orbitals in \ce{UO$_{2}$Cl$_{4}^{2-}$} and approximate models. The orbitals correspond to the bare uranyl and serve as a guide for the eye. Figure S5 of the ESI$\dagger$ lists the orbitals involved in the excitation energies from different models.}
        \label{fig:s1-cl}
    \end{figure*}

    \begin{figure*}
       \centering
        \includegraphics[width=\textwidth]{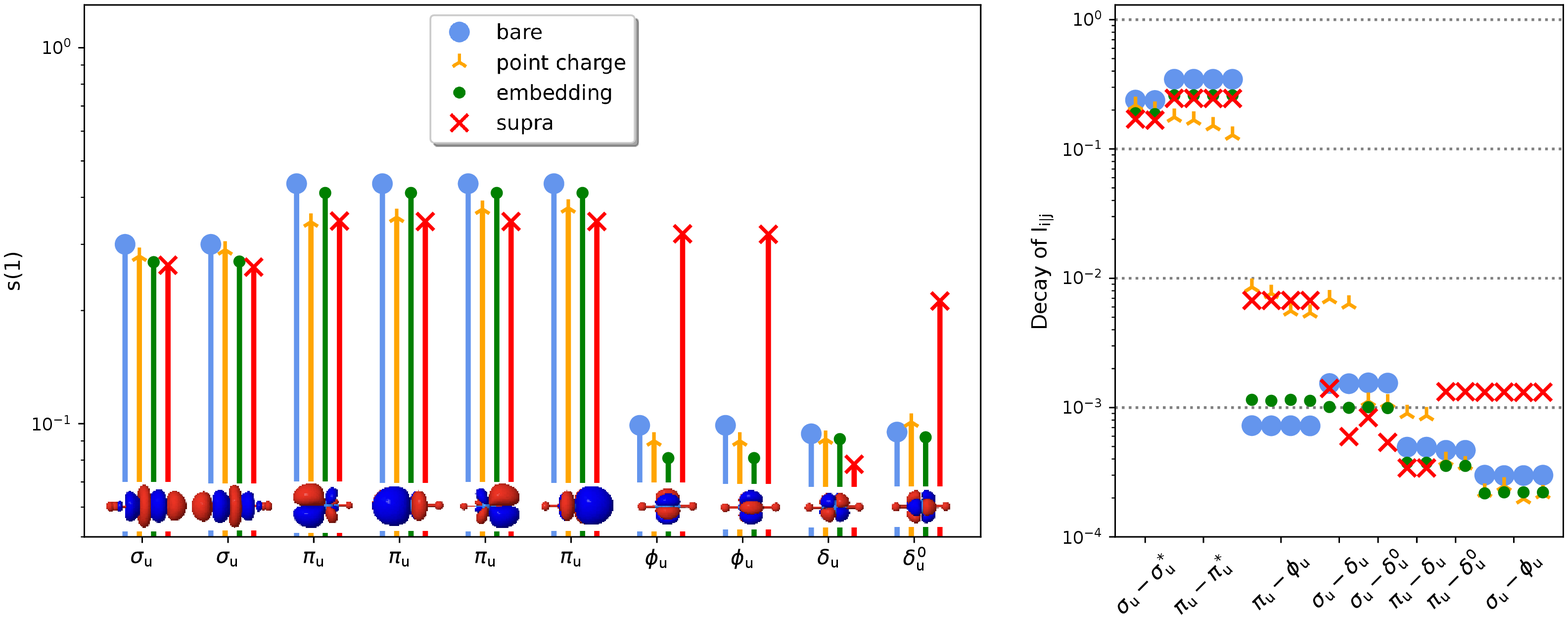}
        \caption{Single orbital entropies (left panel) and decay of the mutual information (right panel) of selected orbitals in \ce{UO$_{2}$Br$_{4}^{2-}$} and approximate models. The orbitals correspond to the bare uranyl and serve as a guide for the eye. Figure S6 of the ESI$\dagger$ lists the orbitals involved in the excitation energies from different models.}
        \label{fig:s1-br}
    \end{figure*}
Figures~\ref{fig:s1-f}--\ref{fig:s1-br} display the single orbital entropies and the decay of the orbital-pair mutual information for the orbitals involved in the lowest-lying excitation energies in the uranyl tetrahalide complexes and their simplified models.
The individual values of the single-orbital entropies are collected in Table S4 of the ESI$\dagger$ and depicted with the corresponding molecular orbital in Figures S1-S3 of the ESI$\dagger$. 
The single-orbital entropies in the left panels of Figures~\ref{fig:s1-f}--\ref{fig:s1-br} demonstrate that the approximate models perform well for the occupied $\sigma_u$ and $\pi_u$ orbitals, but deteriorate from the supramolecular data for the $\phi_u$, $\delta_u$, and $\delta_u^{\prime}$ molecular orbitals.
An exception is the \ce{UO$_{2}$F$_{4}^{2-}$} system, for which the point charge model performs erroneously.
This behavior substantiates the overall bad performance of this model in predicting excitation energies as summarized in Table~\ref{tab:uranyls}.
Interestingly, all supramolecular uranyl tetrahalide structures predict an increased entanglement of $\phi_u$ and $\delta_u^{\prime}$ orbitals, which are not captured by any of the approximate models.
The strong entanglement of the $\delta_u^{\prime}$ orbital in the supramolecular complex is to be expected as non-negligible interactions of that orbital with the environment have been noticed by other authors.~\cite{ruiperez-uo2cl4-jpca-2010, uo2x4-eom-ccsd-ijqc-2022}
At the same time, the corresponding $\delta_u$ orbital is significantly less entangled.

The right panel of Figures~\ref{fig:s1-f}--\ref{fig:s1-br} depicts the mutual information between selected (ground-state) orbitals involved in the lower-lying electronic transitions of the investigated excited states.
The analysis of the $I_{i|j}$ values allows us to monitor how the embedded models can reproduce orbital correlations.
Comparing the mutual information for various embedding schemes with the supramolecular spectrum further allows us to quantify the environmental effects on the system: the larger the deviations in the correlation spectrum between the bare and the supramolecular compounds, the stronger the environmental effects on the orbitals in question.
Figures~\ref{fig:s1-f}--\ref{fig:s1-br} highlight that the WFT-in-DFT embedding schemes correct the orbital pair correlations compared to the bare uranyl and bring them closer to the supramolecular results. 
The selected $I_{i|j}$ spectrum suggests that WFT-in-DFT performs best for the \ce{UO$_{2}$Cl$_{4}^{2-}$} complex, followed by the \ce{UO$_{2}$F$_{4}^{2-}$} compound, and finally \ce{UO$_{2}$B$_{4}^{2-}$}.
Our simple WFT-in-pc model is too restrictive to even qualitatively reproduce the (selected) orbital correlation spectra, which points to the limitations of the point-charge approach.
Especially problematic for embedding models to correct are orbital correlations involving $\delta_u^\prime$ and $\phi_u$ orbitals as their supramolecular counterparts feature correlations with environmental orbitals.
That suggests the good performance of WFT-in-DFT in modeling excited states of \ce{UO$_{2}$F$_{4}^{2-}$} could be just a coincidence or originate from the cancellation of errors in the corresponding excited states.
Note that our implementation allows us to analyze orbital correlations for electronic ground states.
Finally, we should mention that the orbital pair correlations for the \ce{UO2^{2+}}-in-pc model of \ce{UO$_{2}$F$_{4}^{2-}$} are not shown in Figure~\ref{fig:s1-f} due to numerical problems.

\section{Conclusions}\label{sec:conclusions}
In this work, we have implemented a static WFT-in-DFT embedding scheme and a WFT-in-point-charges model, where the WFT part is approximated by pCCD-based methods.
Such an embedding allows us to mimic the effect of the environment with the pCCD family of methods. 
We focused on the EOM-pCCD-LCCSD-in-DFT approach to model electronic excitation energies in the water-ammonia complex and uranyl tetrahalides series of compounds.
Our EOM-pCCD-LCCSD-in-DFT results are comparable to other WFT-in-DFT studies available in the literature but have the advantages of pCCD-based methods to model strong correlation~\cite{oo-ap1rog} and provide an orbital correlation and entanglement analysis.~\cite{kasia-orbital-entanglement-ijqc-2015, ijqc-eratum, post-pccd-entanglement} 
The latter allows us to assess the quality of embedding models (point charges and static DFT embedding) with respect to the supramolecular reference.
Our analysis of single-orbital entropies and the orbital-pair mutual information provides valuable information on orbital entanglement and 
correlation in approximate models compared to supramolecular structures.
Specifically, we showed that the \ce{UO2^{2+}}-in-pc model does not provide a reliable description of orbital entanglement and correlation effects.
That model is particularly erratic for the fluoride environment, reflecting the poor description of excited states. 
The \ce{UO2^{2+}}-in-DFT is less erratic but still has some problems with the accurate description of the uranyl $\Phi_u$ and $\Delta_u$ orbitals in terms of single-orbital entropies and mutual information.
For heavier halides (Cl and Br), we also observe some ligand-to-metal-charge-transfer contributions to the $\pi_u\rightarrow\delta_u$ and $\pi_u\rightarrow\delta_u^{\prime}$ transitions, which pose an additional challenge to embedding models.

To conclude, with the new embedding methods, we can extend the applicability of pCCD-based methods to larger systems, essential, for instance, in organic electronics, and perform a qualitative and quantitative analysis of embedding models in terms of orbital entanglement and correlations.
These orbital entanglement and correlation measures can also be used to define proper fragments of supramolecular structures that minimize the quantum entanglement and correlation of system and environment.

\section{Acknowledgments}\label{sec:acknowledgement}
R.C. and P.T.~acknowledge financial support from the OPUS research grant from the National Science Centre, Poland (Grant No. 2019/33/B/ST4/02114). P.T.~acknowledges the scholarship for outstanding young scientists from the Ministry of Science and Higher Education.
K.B.~acknowledges financial support from a SONATA BIS grant of the National Science Centre, Poland (no.~2015/18/E/ST4/00584).

The authors thank Christoph Jacob for assistance with the PyADF software package.
We have had many helpful discussions with Artur Nowak and Aleksandra Leszczyk.

\nocite{}
\bibliography{stat_emb}

\end{document}